\documentclass{aa}

\usepackage{pifont}
\usepackage{amsmath}
\usepackage{graphicx}
\usepackage{txfonts}
\usepackage{url}
\usepackage{subfigure}
\usepackage{longtable}
\usepackage{lscape}
\usepackage{natbib}
\usepackage[colorlinks,linkcolor=red,anchorcolor=green,citecolor=blue]{hyperref}
\usepackage[normalem]{ulem}

\newcommand{\mum}{\rm {\mu}m}

\newcommand{\nh}{n_{\rm H_2}}
\newcommand{\Tkin}{T_{\rm kin}}

\newcommand{\HII}{H {\small{II}} }
\newcommand{\kms}{{\rm km~s}^{-1}}

\newcommand{\coa}{\rm CO\,(3-2)}
\newcommand{\cob}{\rm CO\,(2-1)}
\newcommand{\coc}{\rm CO\,(1-0)}
\newcommand{\cod}{^{12}{\rm CO}\,{J=3-2}}
\newcommand{\coe}{^{12}{\rm CO}\,{J=2-1}}

\newcommand{\dv}{{\delta}{\rm v}}

\begin{document}

   \title{Using CO line ratios to trace compressed areas in bubble N131}
    \authorrunning{C.-P. Zhang et al.}
    \titlerunning{Infrared dust bubble N131}

    \author{
    Chuan-Peng Zhang\inst{1,2,5}
    \and
    Guang-Xing Li\inst{3}
    \and
    Chenlin Zhou\inst{1,4}
    \and
    Lixia Yuan\inst{1,4}
    \and
    Ming Zhu\inst{1,5}
    }

    \institute{
    National Astronomical Observatories, Chinese Academy of Sciences, 100101 Beijing, P.R. China\\
    \email{cpzhang@nao.cas.cn}
    \and
    Max-Planck-Institut f\"ur Astronomie, K\"onigstuhl 17, D-69117 Heidelberg, Germany
    \and
    South-Western Institute for Astronomy Research, Yunnan University, Kunming, 650500 Yunnan, P.R. China\\
    \email{gxli@ynu.edu.cn}
    \and
    University of Chinese Academy of Sciences, 100049 Beijing, P.R. China
    \and
    CAS Key Laboratory of FAST, National Astronomical Observatories, Chinese Academy of Sciences, 100101 Beijing, P.R. China
    }



  \abstract
   {}
   {N131 is a typical infrared dust bubble showing an expanding ring-like shell. We study the CO line ratios that can be used to trace the interaction in the expanding bubble.}
   {We carried out new $\coa$ observations toward bubble N131 using the 15m JCMT, and derived line ratios by combining these observations with our previous $\cob$ and $\coc$ data from IRAM 30m observations. To trace the interaction between the molecular gas and the ionized gas in the HII region, we used \texttt{RADEX} to model the dependence of the CO line ratios on kinetic temperature and H$_2$ volume density, and examined the abnormal line ratios based on other simulations.}
   {We present $\coa$, $\cob$, and $\coc$ integrated intensity maps convolved to the same angular resolution (22.5$''$). The three different CO transition maps show a similar morphology. The line ratios of $W_{\coa}$/$W_{\cob}$ mostly range from 0.2 to 1.2 with a median of $0.54\pm0.12$, while the line ratios of $W_{\cob}$/$W_{\coc}$ range from 0.5 to 1.6 with a median of $0.84\pm0.15$. The high CO line ratios $W_{\coa}$/$W_{\cob}\gtrsim 0.8 $ and $W_{\cob}$/$W_{\coc}\gtrsim 1.2$ are beyond the threshold predicted by numerical simulations based on the assumed density-temperature structure for the inner rims of the ring-like shell, where the compressed areas are located in bubble N131.}
   {These high CO integrated intensity ratios, such as $W_{\coa}$/$W_{\cob}\gtrsim0.8$ and $W_{\cob}$/$W_{\coc}\gtrsim1.2$, can be used as a tracer of gas-compressed regions with a relatively high temperature and density. This further suggests that the non-Gaussian part of the line-ratio distribution can be used to trace the interaction between the molecular gas and the hot gas in the bubble.}


   \maketitle
%

\section{Introduction}    
\label{sect:intro}

Infrared dust bubbles are ubiquitous interstellar objects \citep{chur2006,chur2007,simp2012,Hou2014,N131_2013,N131_2016,Jayasinghe2019}. However, the details of the bubble shell formation mechanism are still unclear \citep[e.g.,][]{beau2010,wats2008}. N131 is a quite typical bubble, which has been observed and investigated in detail by \citet{N131_2013,N131_2016}. Bubble N131 has an inner minor radius of 13\,pc and an inner major radius of 15\,pc at a kinetic distance of $\sim$8.6\,kpc, and the center coordinates are R.A.(J2000) = $19\rm ^h52\rm ^m21.\!\!^{\rm s}5$, DEC.(J2000) = $+26^{\circ}21'24.\!\!{''}0$. A ring-like shell is visible at 8.0 and 24\,$\mum$ and is associated with CO emission (see Figure\,\ref{Fig:co_int}). Two giant elongated molecular clouds are located at opposite sides of the ring-like shell, and together, they exhibit a large velocity gradient. In addition, there is a huge cavity inside the bubble that is visible in the $5.8 - 500\,\mum$ emission. The column density, excitation temperature, and velocity of the $\coc$ emission show a possibly stratified structure from the inner to outer rims of the ring-like shell. These suggest that bubble N131 has an expanding shell caused by feedback of strong stellar winds from the star formation at the center of the bubble \citep[see also the detailed discussion in][]{N131_2016}.

The $\coa$, $\cob$, and $\coc$ transitions have different upper energy levels \citep{Kaufman1999}. The different transitions can therefore be used to trace different excitation conditions. The integrated intensity ratios, such as $W_{\coa}$/$W_{\cob}$ and $W_{\cob}$/$W_{\coc}$, may indicate a different temperature and density structure of the molecular cloud environments \citep{Hasegawa1994,Wilson1997}. For example, high $W_{\cob}/W_{\coc}$ ratios have been observed in the Large Magellanic Cloud (LMC) by \citet{Bolatto2000}. It was proposed that self-absorbed emission and optical depth effects may be possible origins for the high line ratios \citep{Bolatto2000,Bolatto2003}. Additionally, the line ratios are also quite important for us to diagnose the evolutionary stage of the molecular clouds \citep[e.g.,][]{Sakamoto1995, Beuther2000,Yoda2010,Polychroni2012,Nishimura2015}.

In this work, we carry out new $\coa$ observations toward bubble N131 using the 15m James Clerk Maxwell Telescope (JCMT). In combination with our previous  $\cob$ and $\coc$ line observations with the IRAM 30m telescope, we study how the CO line ratios can be used to trace the interaction in the expanding infrared dust bubble N131. In Section\,\ref{sect:obser} we describe the observations and data reduction. In Section\,\ref{sect:result_analysis} we show the observational results and the \texttt{RADEX} modeling. In Section\,\ref{sec:discussion} we mainly discuss the possibility  of using the CO line ratios to trace the compressed inner rims of the ring-like shell around the bubble. In Section\,\ref{sect:summary} we summarize our results.


\section{Observations}
\label{sect:obser}

\subsection{$\cod$}
\label{sect:co32}

We carried out new $\coa$ observations (M17BP077 and M18BP069) toward bubble N131 during September 2017 -- August 2018 using the Heterodyne Array Receiver Programme \citep[HARP;][]{Buckle2009} at the 15m JCMT. Maps were referenced against an off-source position that was free of any significant CO emission in the \citet{Dame2001} CO Galactic Plane Survey. At 345\,GHz, the half-power beam width (HPBW) was $\sim$14.0$''$, and the main beam efficiency is $\eta_{\rm mb}=0.64$, taken from the JCMT efficiency archive. The main beam brightness temperature ($T_{\rm mb}$) can be derived by $T_{\rm mb} = T^*_{\rm A}/\eta_{\rm mb}$. The on-the-fly mapping mode was used to scan the bubble with a sampling step of 7.0$''$. For further line ratio analysis, the raw data were then convolved to the same angular resolution of 22.5$''$, corresponding to the lowest angular resolution of $\coc$ (see Section\,\ref{sect:co21_obs}), with a grid of 11.0$''$ using the \texttt{GILDAS}\footnote{\url{http://www.iram.fr/IRAMFR/GILDAS/}} software package.

Calibration scans, pointing, and focus were performed regularly. Calibration scans were taken at the beginning of each subscan. A pointing was made about every hour. A focus scan was taken every three hours, but more scans were taken around sunset and sunrise. The flux calibration is expected to be accurate to within 10\%. The \texttt{GILDAS} software package was used to reduce the observational data.

\subsection{$\coe$ and $J=1-0$}
\label{sect:co21_obs}

Our $\cob$ and $\coc$ observations were simultaneously carried out in April 2014 using the IRAM 30m telescope\footnote{Based on observations carried out with the IRAM 30m Telescope. IRAM is supported by INSU/CNRS (France), MPG (Germany), and IGN (Spain).} on Pico Veleta, Spain. The observations have been introduced in detail in our previous work in \citet{N131_2016}. In our raw data, the HPBWs of $\cob$ and $\coc$ are 11.3$''$ and 22.5$''$ , respectively, with the same sampling step of 9.3$''$. For further line ratio analysis, the raw data were then convolved to the lowest angular resolution of 22.5$''$ with a grid of 11.0$''$ using the \texttt{GILDAS} software package.

\section{Results and analysis}
\label{sect:result_analysis}

\subsection{CO integrated intensity distributions}
\label{sect:co_distri}

\begin{figure}
\centering
\includegraphics[width=0.49\textwidth, angle=0]{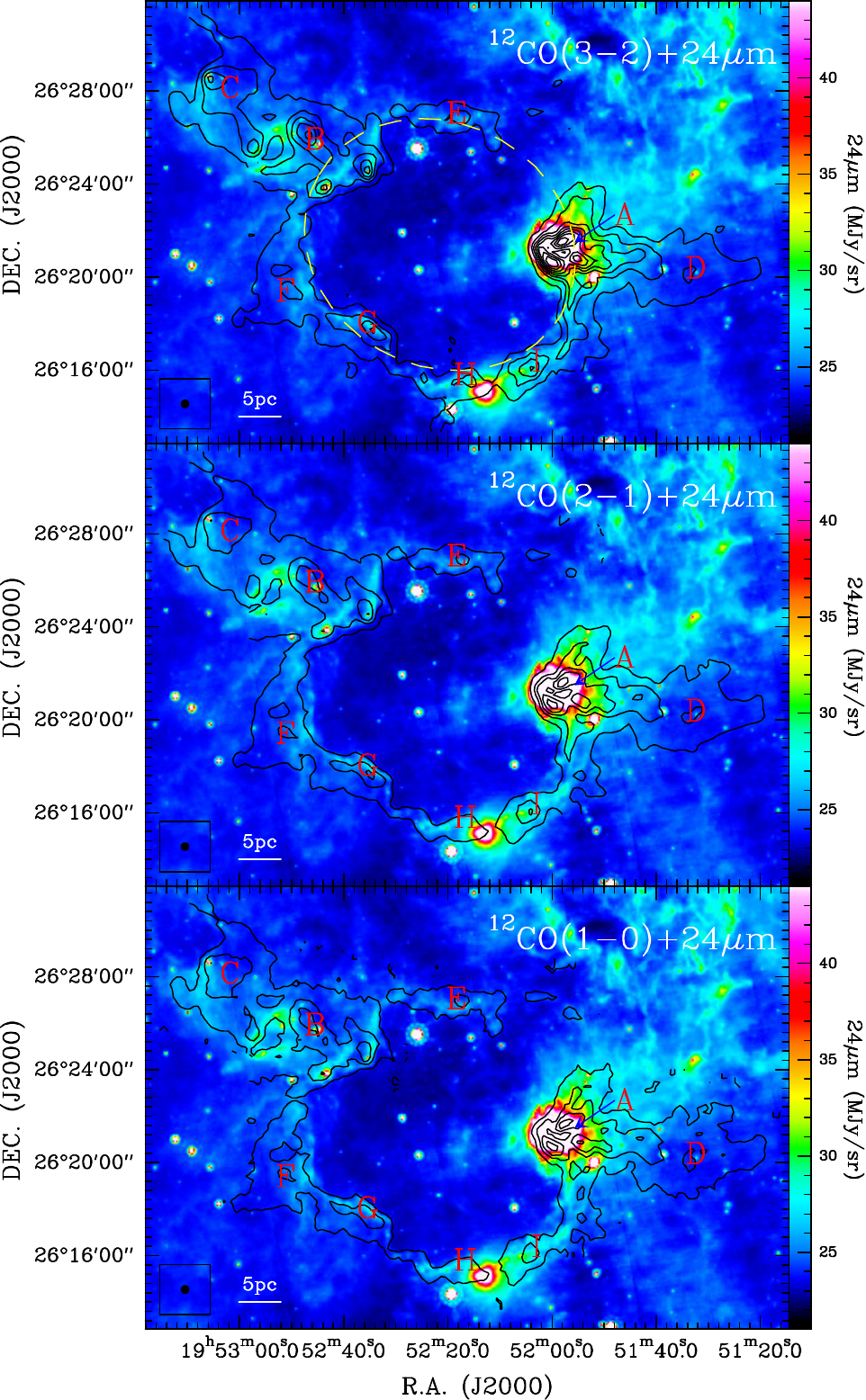}
\caption{Integrated intensity maps of $\coa$ (\textit{upper}), $\cob$ (\textit{middle}), and $\coc$ (\textit{lower}) lines with a velocity range from $-16.0$ to $-5.0\,\kms$ superimposed on 24\,$\mum$ emission. The contour levels in each CO map start at 5$\sigma$ in steps of 10$\sigma$ with $\sigma_{\coa} = 0.6\,{\rm K}\,\kms$, $\sigma_{\cob} = 1.3\,{\rm K}\,\kms$, and $\sigma_{\coc} = 1.6\,{\rm K}\,\kms$. The letters and the ellipse indicate the positions of nine molecular clumps (A-I) and the ring-like shell of the bubble, respectively. The angular resolution (22.5$''$) is indicated in the bottom left corner of each panel.}
\label{Fig:co_int}
\end{figure}

Figure\,\ref{Fig:co_int} displays the integrated intensity maps of $\coa$, $\cob$, and $\coc$ lines with a velocity range from $-16.0$ to $-5.0\,\kms$ superimposed on MIPSGAL 24\,$\mum$ emission \citep{care2009}. All the CO data were convolved to the same angular resolution of 22.5$''$. We also label the nine selected molecular clumps \citep{N131_2016} and the ring-like shell of the bubble in the maps. The morphological structures of the three integrated intensity maps are clearly similar. 

\subsection{Spectra}
\label{sect:spectra}

\begin{figure}
\centering
\includegraphics[width=0.49\textwidth, angle=0]{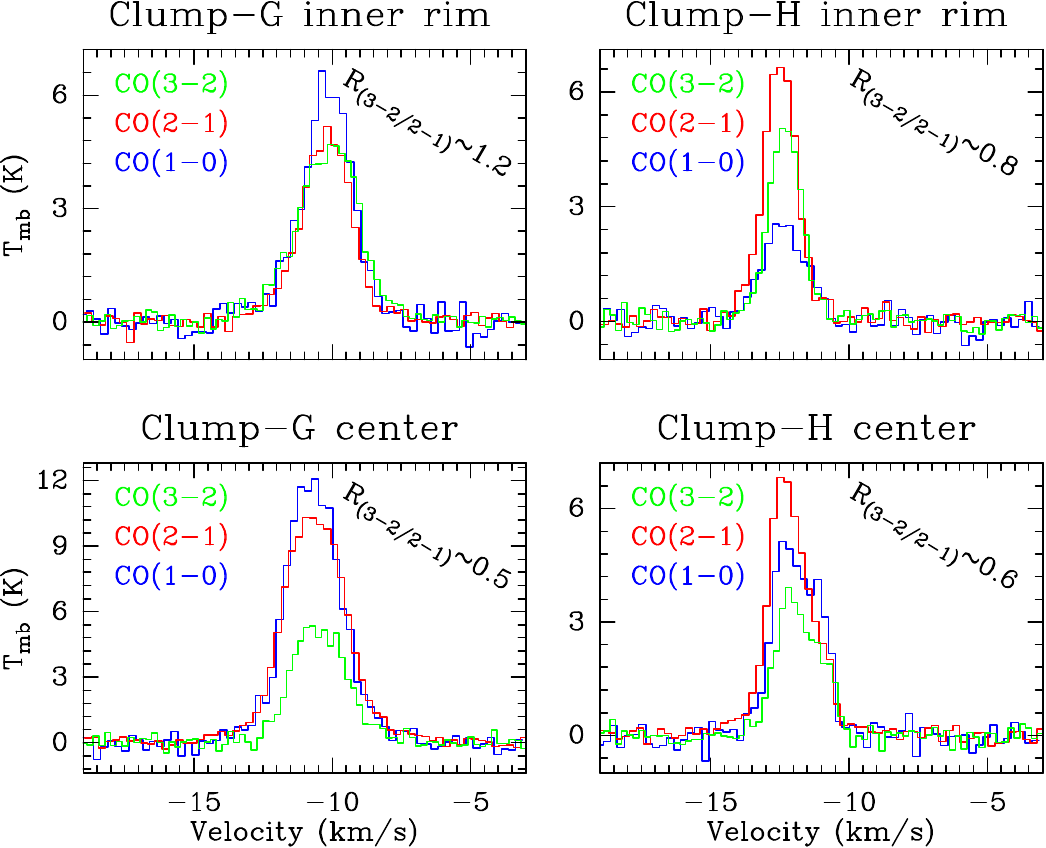}
\caption{Example spectra of the high ratios $W_{\coa}$/$W_{\cob}$ (\textit{upper}) from the inner rims of the ring-like shell near clumps\,G and H, and of the low ratios (\textit{lower}) from the clump center regions (see also Figure\,\ref{Fig:ratio_int}).}
\label{Fig:spectra3221}
\end{figure}

\begin{figure}
\centering
\includegraphics[width=0.49\textwidth, angle=0]{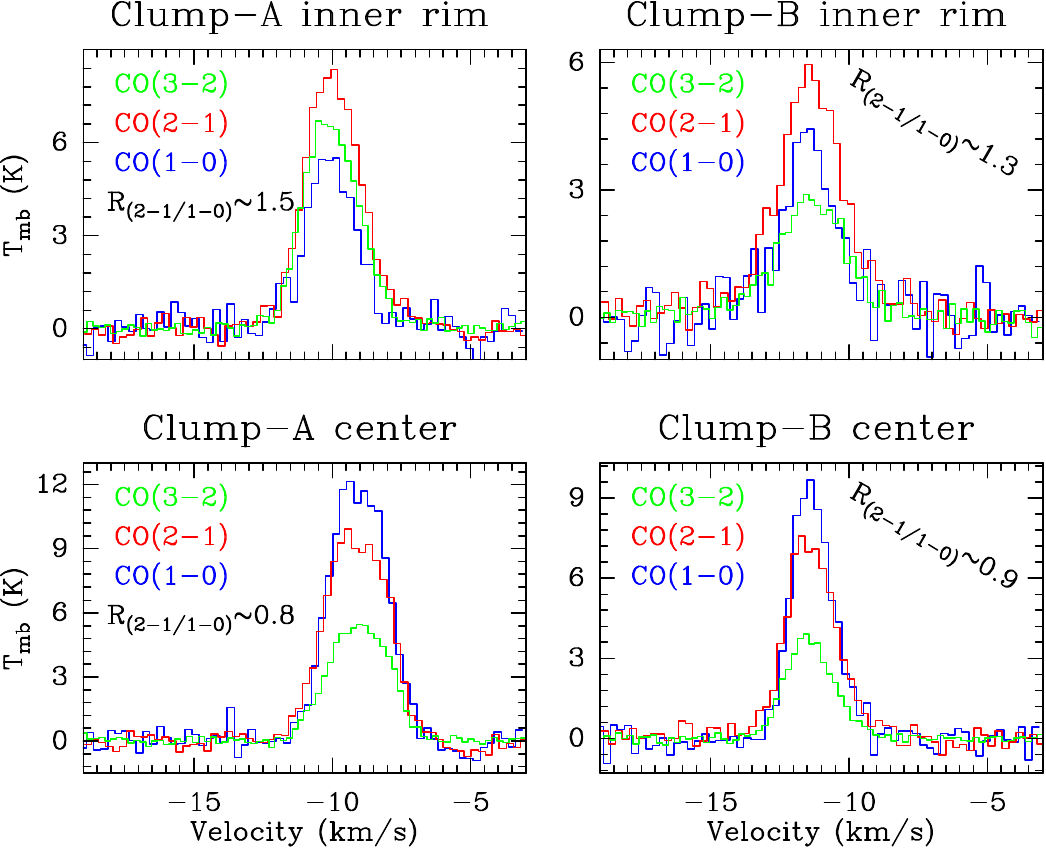}
\caption{Example spectra of the high ratios $W_{\cob}$/$W_{\coc}$ (\textit{upper}) from the inner rims of the ring-like shell near clumps\,A and B, and of the low ratios (\textit{lower}) from the clump center regions (see also Figure\,\ref{Fig:ratio_int}).}
\label{Fig:spectra2110}
\end{figure}

We extracted several example spectra $\coa$, $\cob$, and $\coc$ (see upper panels in Figures\,\ref{Fig:spectra3221} and \ref{Fig:spectra2110}) with high ratios ($W_{\coa}$/$W_{\cob}\gtrsim0.8$ and $W_{\cob}$/$W_{\coc}\gtrsim1.2$) from the inner rims near clumps\,A, B, G, and H. All the spectra with the highest ratios have high signal-to-noise ratios above 10$\sigma$. This indicates that the line ratios have high signal-to-noise ratios at least above 7$\sigma$. For comparison, we also extracted some spectra (see lower panels in Figure\,\ref{Fig:spectra3221} and \ref{Fig:spectra2110}) with low line ratios from the corresponding clump center regions. 

\subsection{Integrated intensity ratio distributions}
\label{sect:int_ratio_distri}

\begin{figure}
\centering
\includegraphics[width=0.49\textwidth, angle=0]{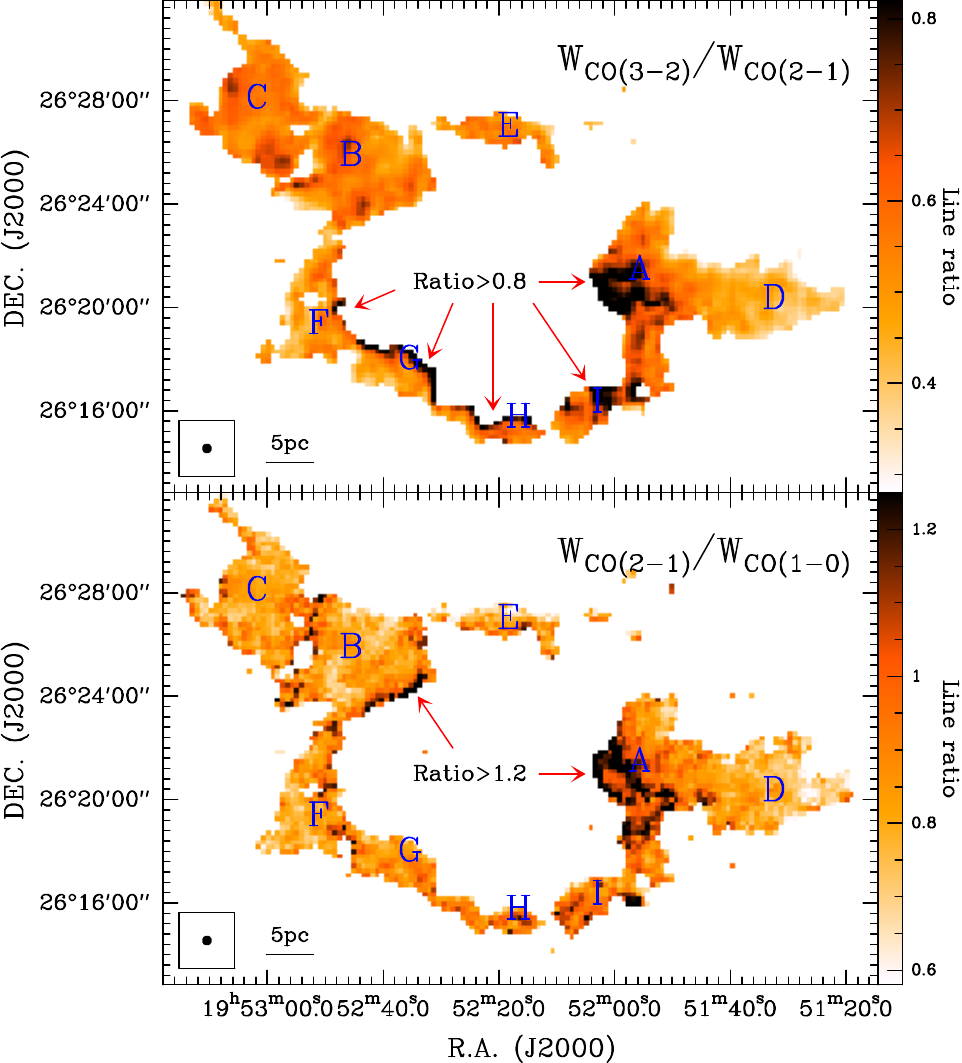}
\caption{Integrated intensity ratio maps of $W_{\coa}$/$W_{\cob}$ (\textit{upper}) and $W_{\cob}$/$W_{\coc}$ (\textit{lower}) derived from the integrated intensity maps that are above 5$\sigma$ in Figure\,\ref{Fig:co_int}. The letters indicate the positions of nine molecular clumps (A-I) in the bubble. The angular resolution is indicated in the bottom left corner of each panel.}
\label{Fig:ratio_int}
\end{figure}

\begin{figure}
\centering
\includegraphics[width=0.45\textwidth, angle=0]{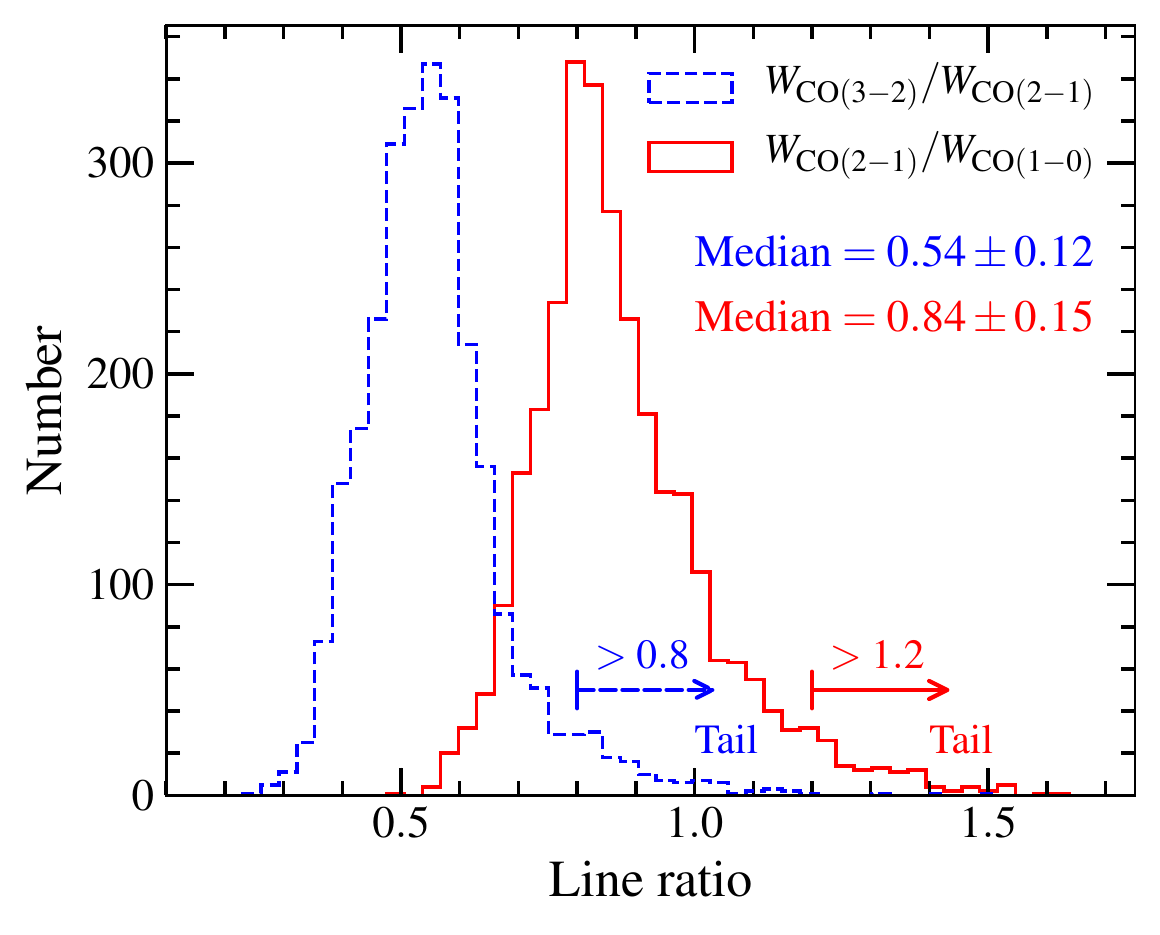}
\caption{Integrated intensity ratio histograms of $W_{\coa}$/$W_{\cob}$ and $W_{\cob}$/$W_{\coc}$ for all pixels in Figure\,\ref{Fig:ratio_int}. The median uncertainties are derived from the standard deviation of the sample.}
\label{Fig:hist_ratio_int}
\end{figure}

Figure\,\ref{Fig:ratio_int} displays the integrated intensity ratio maps of $W_{\coa}$/$W_{\cob}$ and $W_{\cob}$/$W_{\coc}$. The ratios were obtained based on the integrated intensity maps that are above 5$\sigma$ (see Figure\,\ref{Fig:co_int}). For the line ratios we considered pixels above $3.5\sigma$ according to the error propagation of the integrated intensity maps. It clearly shows that at clumps A, F, G, H, and I in the $W_{\coa}$/$W_{\cob}$ map, the inner rims of ring-like shell have a higher integrated intensity ratio ($W_{\coa}$/$W_{\cob}\gtrsim0.8$) than the outer rims, while in the $W_{\cob}$/$W_{\coc}$ map the highest line ratio occurs at the inner rims of the shell near clumps A, B, E, and F with $W_{\cob}$/$W_{\coc}\gtrsim1.2$. Figures\,\ref{Fig:spectra3221} and \ref{Fig:spectra2110} display some spectra $\coa$, $\cob$, and $\coc$ extracted from the inner rims of the ring-like shell near clumps\,A, B, G, and H with high ratios ($W_{\coa}$/$W_{\cob}\gtrsim0.8$ and $W_{\cob}$/$W_{\coc}\gtrsim1.2$).

Figure\,\ref{Fig:hist_ratio_int} displays the integrated intensity ratio histograms of $W_{\coa}$/$W_{\cob}$ and $W_{\cob}$/$W_{\coc}$ for all pixels in Figure\,\ref{Fig:ratio_int}. The line ratios of $W_{\coa}$/$W_{\cob}$ mostly range from 0.2 to 1.2 with a median of $0.54\pm0.12$, which is slightly lower than what was found ($\approx$0.75) at the Central Molecular Zone of the Milky Way \citep{Kudo2011}. The line ratios of $W_{\cob}$/$W_{\coc}$ range from 0.5 to 1.6 with a median of $0.84\pm0.15$. We also derived the median value of $W_{\coa}$/$W_{\coc}$ , which is around 0.45, close to the average value of $W_{\coa}$/$W_{\coc}\approx0.5$ in star-forming galaxies \citep[e.g,][]{Aravena2010,Aravena2014,Daddi2015}.

\subsection{\texttt{RADEX} modeling}
\label{sect:radex}

To study the line ratio distributions as a function of kinetic temperature and H$_2$ volume density in bubble N131, we used the nonlocal thermodynamic equilibrium (non‐LTE) radiative transfer code \texttt{RADEX}\footnote{\url{https://home.strw.leidenuniv.nl/~moldata/radex.html}} \citep{Tak2007} with the Leiden Atomic and Molecular Database \citep[LAMDA;][]{Schoier2005} to model the $\coa$, $\cob$, and $\coc$ lines. The model grid extends over a grid of 51 temperatures ($\Tkin=3-500$\,K) and 51 densities ($\nh=10-10^{5}\,\rm cm^{-3}$). The CO column density and line width were fixed with $N_{\rm CO}=2.2\times10^{17}\,\rm cm^{-2}$ and $\dv=3.5\,\kms$, which are the derived median values of CO column density and $\coc$ velocity dispersion from CO\,($1-0$) and $^{13}$CO\,($1-0$) in N131 \citep[see][]{N131_2016}. The beam-filling factors were assumed to be unity.

Figures\,\ref{Fig:temperature_density_3221} and \ref{Fig:temperature_density_2110} display the line ratio and optical depth distributions as a function of kinetic temperature and H$_2$ volume density obtained with \texttt{RADEX} modeling. Linear molecules of CO at low rotational transitions (critical density of about $n_{\rm crit}\sim10^4\rm\,cm^{-3}$) are tracers of low-density gas \citep{Kaufman1999,Qin2008,Nishimura2015,Penaloza2018}. For a given molecule, moving up to a high rotational transition will lead to a high critical density. The high rotational transitions are sensitive to a high temperature based on the large velocity gradient (LVG) model. The high temperature and density can therefore be probed with the high CO line ratios \citep{Tak2007}. 



\section{Discussion: Line ratios tracing the compressed areas}
\label{sec:discussion}


\citet{Wilson1997} found that the $W_{\coa}$/$W_{\cob}$ line ratios for the molecular clouds containing optical \HII regions ($0.79\pm0.05$) are somewhat higher than those for the clouds without optical \HII regions ($0.58\pm0.06$), while the line ratio in the giant \HII region is even higher ($1.07\pm0.03$). \citet{Wilson1997} also suggested that the high line ratio may be caused by heating of the gas by the massive stars. Line ratio distributions such as $W_{\coa}$/$W_{\cob}$ and $W_{\cob}$/$W_{\coc}$ have been used to study the interaction in supernova remnant molecular cloud system \citep[e.g,][]{Jiang2010,Zhou2016,Zhou2018,Arias2019}. The high ratios with $W_{\cob}$/$W_{\coc}\approx1.6$ were suggested by \citet{Zhou2016} to trace the shocked compressed gas that is located at the shell of supernova remnant Tycho. Recently, \citet{Celis2019} also found that the high integrated line ratios $W_{\coa}$/$W_{\cob}$ at the shell of the LMC supergiant bubble N11 may be caused by the expansion of nebulae and the interaction with radiation from OB association. The question now is why and how the CO line ratios can be used to trace the interactions.

The infrared dust bubble N131 originates from expanding \HII regions, but the \HII region inside has been extinguished \citep{N131_2013,N131_2016}. Figure\,\ref{Fig:ratio_int} clearly shows that most parts of the inner rims of the ring-like shell have higher integrated intensity ratios (e.g., $W_{\coa}$/$W_{\cob}\gtrsim0.8$, $W_{\cob}$/$W_{\coc}\gtrsim1.2$) than the outer rims. Additionally, the most notable discrepancy between the two ratio distributions is that at the inner rims of the ring-like shell near clumps G and H, the ratio $W_{\coa}$/$W_{\cob}$ is much higher than in other regions (except for the complicated clump A\footnote{Clump A is a small expanding \HII region that is deeply embedded in the ring-like shell of bubble N131 \citep[see details in][]{N131_2016}.}) but the ratio $W_{\cob}$/$W_{\coc}$ is not, while at the inner rims near clump B, the ratio $W_{\cob}$/$W_{\coc}$ is much higher than in other regions but the ratio $W_{\coa}$/$W_{\cob}$ is not. This may suggest that the inner rims of the ring-like shell near clumps G and H have a relatively high kinetic temperature up to the excitation temperature of high transition level of $\coa$, leading to stronger $\coa$ emission than in other regions; while the inner rims near clump B have a relatively low kinetic temperature just up to the low transition level of $\cob$, leading to stronger $\cob$ emission than in other regions. This also suggests that the inner rims of the ring-like shell were compressed by strong stellar winds from the bubble insides \citep[see also discussion in][]{Nishimura2015}.

\begin{figure}
\centering
\includegraphics[width=0.45\textwidth, angle=0]{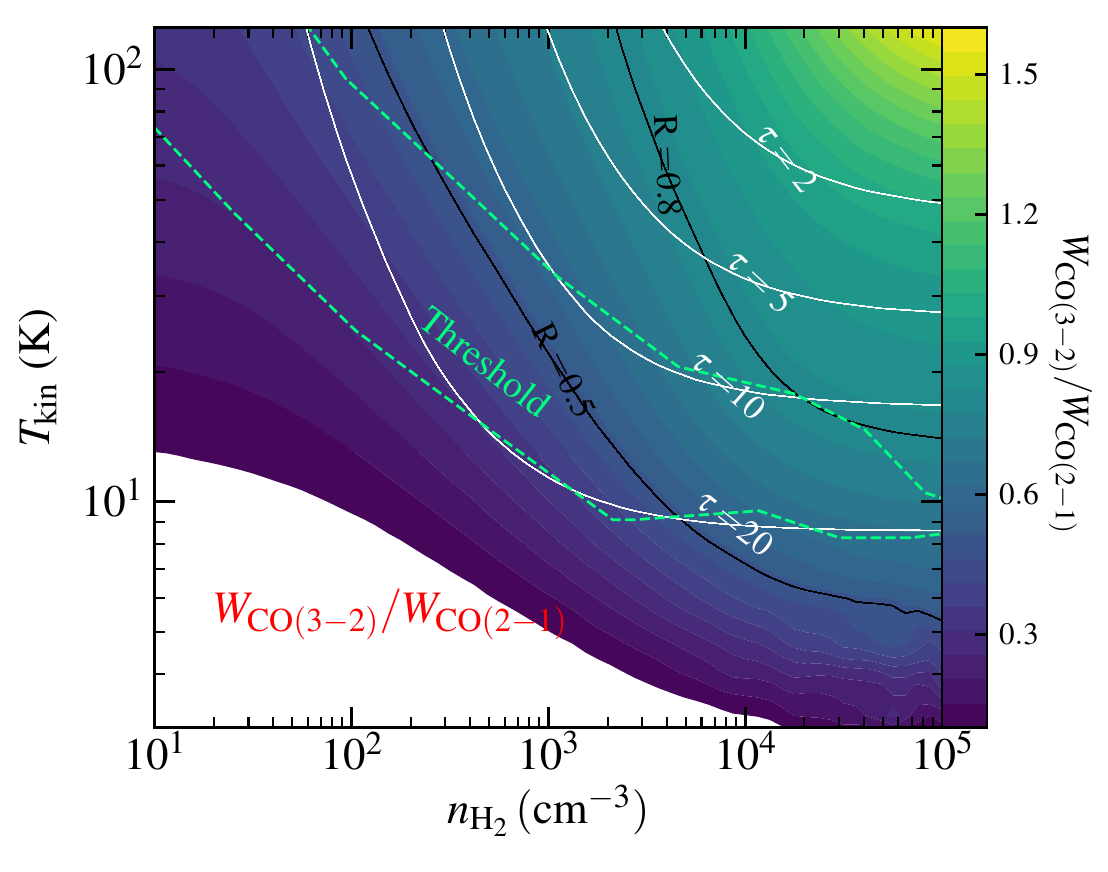}
\caption{Line ratios ($R=W_{\coa}$/$W_{\cob}$) and optical depths ($\tau_{\cob}$) in the conditions of $N_{\rm CO}=2.2\times10^{17}\,\rm cm^{-2}$ and $\dv=3.5\,\kms$ (estimated by median values in N131) as a function of kinetic temperature and volume density by \texttt{RADEX} modeling. The green contour indicates a region (or threshold) for a possible gas temperature-density distribution in a colliding flow at the onset of star formation from simulations in \citet{Clark2012}.}
\label{Fig:temperature_density_3221}
\end{figure}

\begin{figure}
\centering
\includegraphics[width=0.45\textwidth, angle=0]{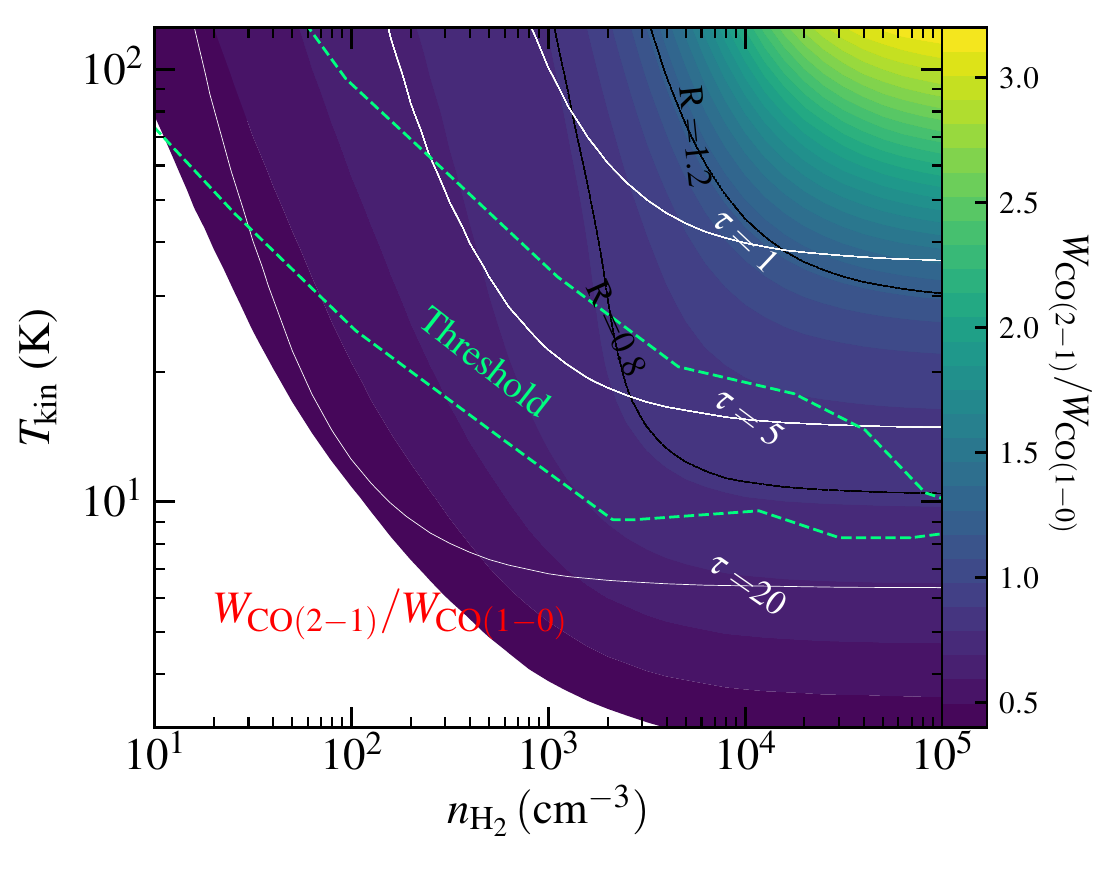}
\caption{Line ratios ($R=W_{\cob}$/$W_{\coc}$) and optical depths ($\tau_{\coc}$) in the conditions of $N_{\rm CO}=2.2\times10^{17}\,\rm cm^{-2}$ and $\dv=3.5\,\kms$ (estimated by median values in N131) as a function of kinetic temperature and volume density by \texttt{RADEX} modeling. The green contour indicates a region (or threshold) for a possible gas temperature-density distribution in a colliding flow at the onset of star formation from simulations in \citet{Clark2012}.}
\label{Fig:temperature_density_2110}
\end{figure}

To trace the compressed inner rims of the ring-like shell by stellar winds from the bubble insides, we computed the expected CO line ratios at different gas temperatures and densities using \texttt{RADEX} code. The results are presented in Figures\,\ref{Fig:temperature_density_3221} and \ref{Fig:temperature_density_2110}. We then determined the CO line ratios that can be used to trace the interactions. We recall that in an ordinary molecular cloud, the cold gas is mainly heated by cosmic rays. This heating is balanced by radiative cooling \citep{Draine2011}. As a result, we expect a limited range of temperatures and densities for the molecular gas, which leads to a limited range of observed line ratios. These line ratios, which lie far beyond the upper limit, could trace the interaction between the cold and hot gas that presumably lies in the inner rims of a bubble shell because these interactions should increase the temperature and density.

Therefore, we propose to use CO line ratios $W_{\coa}$/$W_{\cob}\gtrsim0.8$ and $W_{\cob}$/$W_{\coc}\gtrsim1.2$ to trace the compressed inner rims of the ring-like shell. The thresholds were selected based on the following considerations. The thresholds correspond to the non-Gaussian tail of line ratio distribution presented in Figure\,\ref{Fig:hist_ratio_int}, where we propose that non-interacting clouds should produce line ratios that are Gaussian distributed, and the non-Gaussian parts of the distributions are caused by interaction. To justify our thresholds, we used \texttt{RADEX} to compute the line ratios as a function of gas temperature and density (see Figures\,\ref{Fig:temperature_density_3221} and \ref{Fig:temperature_density_2110}). By overlaying the expected range of gas density and temperature found in the most recent numerical simulations\footnote{Although the simulations in \citet{Clark2012} were carried out under a certain set of initial conditions, the predicted temperature-density relation for the molecular gas is relatively robust (e.g., independent of the initial condition) and is applicable to our data. Additionally, due to the short cooling times, the density-temperature relation of the molecular gas should not depend on the initial conditions (e.g., whether the converging speed is fast or slow).} \citep{Clark2012}, we derived the expected CO line ratios for non-interacting clouds. The highest ratios are located in regions with moderate or low optical depths ($\tau\lesssim5$ for $W_{\coa}$/$W_{\cob}\gtrsim0.8$ and $\tau\lesssim1$ for $W_{\cob}$/$W_{\coc}\gtrsim1.2$) in the temperature-density plane. Line ratios higher than this can be used to trace the interaction region where the gas temperature and density are higher than normal.

\section{Summary}
\label{sect:summary}

Based on our previous multiwavelength observations \citep{N131_2013,N131_2016}, the infrared dust bubble N131 is a typical bubble showing an expanding ring-like shell, which has been swept up by the energetic winds of ionizing stars inside. We here carried out new $\coa$ observations toward the bubble N131 using the 15m JCMT, and also used our published $\cob$ and $\coc$ line data observed with the IRAM 30m telescope. We plotted their integrated intensity maps, which were convolved to the same angular resolution (22.5$''$). We find that the three different CO transition maps show a similar morphological structure.

In bubble N131, we used the \texttt{RADEX} code to model the kinetic temperature and H$_2$ volume density, and we studied the relationship between them and line ratios. The line ratios of $W_{\coa}$/$W_{\cob}$ mostly range from 0.2 to 1.2 with a median of $0.54\pm0.12$, while the line ratios of $W_{\cob}$/$W_{\coc}$ range from 0.5 to 1.6 with a median of $0.84\pm0.15$. The line width ratios between $\coa$, $\cob$, and $\coc$ are close to unity.

To probe the interaction between the hot stellar winds and the cold molecular ring-like shell, we performed \texttt{RADEX} modeling to test the dependence of the line ratios on the underlying parameters such as temperature and density, and to predict the range of CO integrated intensity ratios $W_{\coa}$/$W_{\cob}$ and $W_{\cob}$/$W_{\coc}$ if the gas temperatures and densities are predicted by the chemodynamics simulations. Line ratios far beyond the temperature-density threshold \citep{Clark2012} could thus be used to trace the interactions.

From our observations, we find that the high CO integrated intensity ratios $W_{\coa}$/$W_{\cob}$ and $W_{\cob}$/$W_{\coc}$ are far beyond the prediction from the most recent numerical simulation without stellar feedback. As a result, these high line ratios can be used to trace the compressed areas in bubble N131. We suggest that the high CO integrated intensity ratios, such as $W_{\coa}$/$W_{\cob}\gtrsim0.8$ and $W_{\cob}$/$W_{\coc}\gtrsim1.2$, can be used as a tracer of gas-compressed regions with a relatively high temperature and density. We further proved that the non-Gaussian part of the line-ratio distribution can be used to trace the interaction between the molecular gas and the hot gas in the bubble.

\begin{acknowledgements}
We thank the anonymous referees for constructive comments that improved the manuscript. This work is supported by the National Natural Science Foundation of China Nos.\,11703040, 11743007, and National Key Basic Research Program of China (973 Program) No.\,2015CB857101. C.-P. Zhang acknowledges support by the MPG-CAS Joint Doctoral Promotion Program (DPP) and China Scholarship Council (CSC) in Germany as a postdoctoral researcher. The JCMT is operated by the EAO on behalf of NAOJ; ASIAA; KASI; CAMS as well as the National Key R\&D Program of China (No.\,2017YFA0402700). Additional funding support is provided by the STFC and participating universities in the UK and Canada.

\end{acknowledgements}

\bibliographystyle{aa}
\bibliography{references}

\appendix

\section{Line width ratio}
\label{sect:width_ratio_distri}

\begin{figure}
\centering
\includegraphics[width=0.49\textwidth, angle=0]{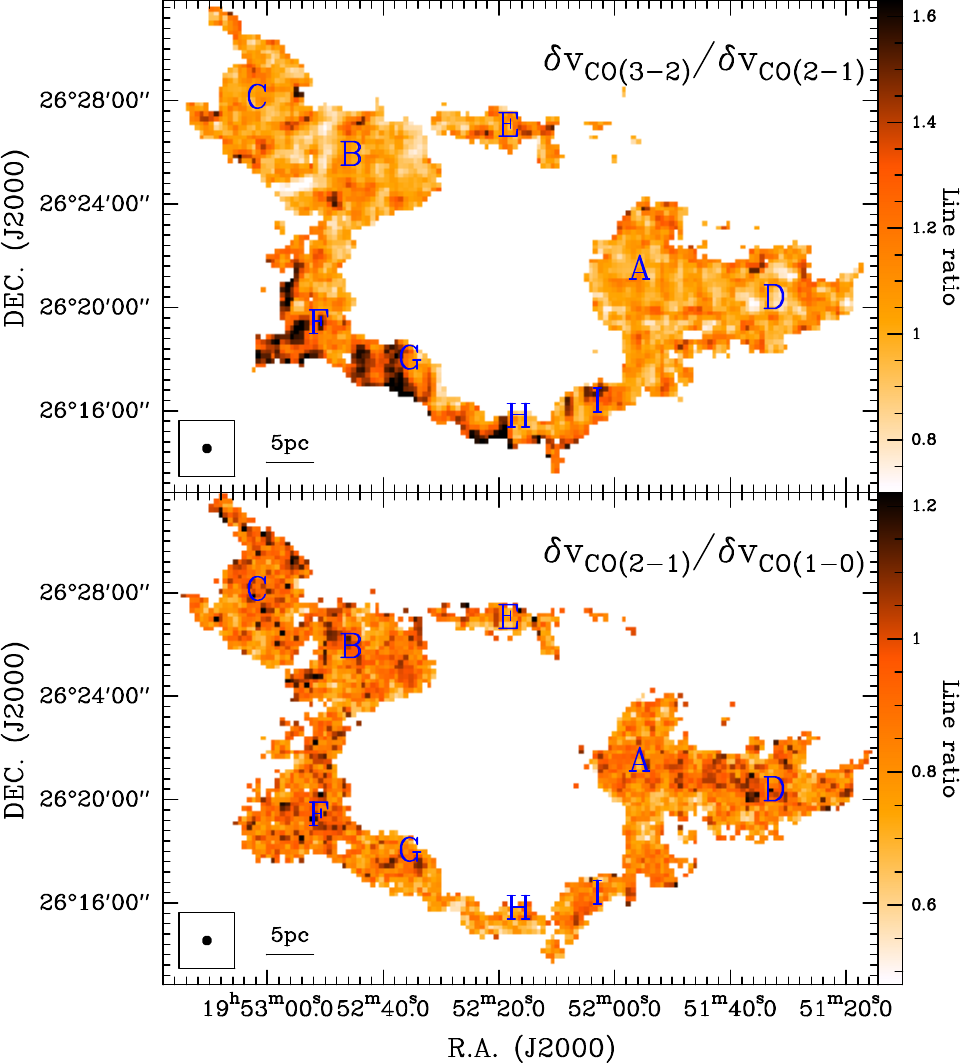}
\caption{Line width ratio maps of $\dv_{\coa}$/$\dv_{\cob}$ (\textit{upper}) and $\dv_{\cob}$/$\dv_{\coc}$ (\textit{lower}). The letters indicate the positions of the nine molecular clumps (A-I) in the bubble. The angular resolution is indicated in the bottom left corner.}
\label{Fig:ratio_width}
\end{figure}

\begin{figure}
\centering
\includegraphics[width=0.45\textwidth, angle=0]{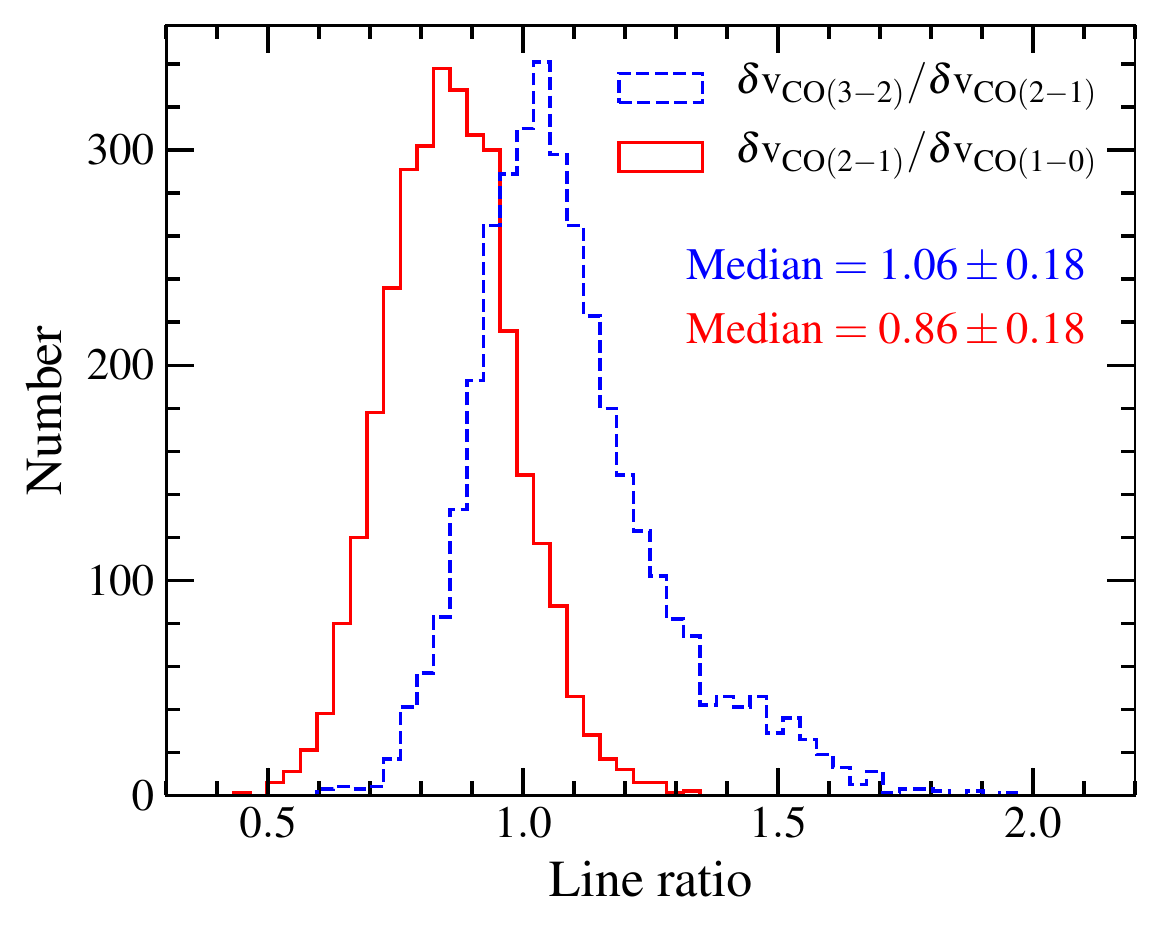}
\caption{Line width ratio histograms of $\dv_{\coa}$/$\dv_{\cob}$ and $\dv_{\cob}$/$\dv_{\coc}$ for all pixels in Figure\,\ref{Fig:ratio_width}. }
\label{Fig:hist_ratio_width}
\end{figure}

Figure\,\ref{Fig:ratio_width} displays the line width ratio maps of $\dv_{\coa}$/$\dv_{\cob}$ and $\dv_{\cob}$/$\dv_{\coc}$. For clumps F, G, and H in the $\dv_{\coa}$/$\dv_{\cob}$ map, the outer rims of the ring-like shell have a higher line width ratio than the inner rims, while this is reversed for clump I. For the other clumps, there is no visible line ratio gradient feature. In the $\dv_{\cob}$/$\dv_{\coc}$ map, it seems that the higher line width ratios are located at the clump center positions, and the line ratio gradient is not evident. Figure\,\ref{Fig:hist_ratio_width} displays the line width ratio histograms of $\dv_{\coa}$/$\dv_{\cob}$ and $\dv_{\cob}$/$\dv_{\coc}$ for all pixels in Figure\,\ref{Fig:ratio_width}. The line ratios of $\dv_{\coa}$/$\dv_{\cob}$ mostly range from 0.6 to 1.8 with a median of $1.06\pm0.18$, while $\dv_{\cob}$/$\dv_{\coc}$ range from 0.5 to 1.3 with a median of $0.86\pm0.18$. We can also derive that the median value of $\dv_{\coa}$/$\dv_{\coc}$ is around 0.91. Compared with the line width ratios, we have $\dv_{\coc}>\dv_{\coa}>\dv_{\cob}$ only for their median values. However, generally, the line width ratios between $\coa$, $\cob$, and $\coc$ are close to unity.

\end{document}